\documentclass[conference]{IEEEtran}

\usepackage{cite}
\usepackage{amsmath,amssymb,amsfonts,amsthm}
\usepackage{algorithmic}
\usepackage{graphicx}
\usepackage{mathtools}
\usepackage{numprint}
\usepackage{textcomp}
\usepackage{url}
\usepackage{xcolor}
\usepackage{xfrac}
\usepackage[sort&compress,numbers]{natbib}

\newtheorem{remark}{Remark}

\ifCLASSINFOpdf
\else
\fi
\begin{document}

%\title{Performance Analysis of Universal Decoders in the Presence of Alpha-Stable Noise Channels}
\title{Error correction in interference-limited wireless systems}

\author{\IEEEauthorblockN{Charles Wiame}
\IEEEauthorblockA{Research Laboratory of Electronics \\
Massachusetts Institute of Technology\\
wiame@mit.edu}
\and
\IEEEauthorblockN{Ken R. Duffy}
\IEEEauthorblockA{College of Engineering and College of Science\\
Northeastern University\\
k.duffy@northeastern.edu}
\and
\IEEEauthorblockN{Muriel Médard}
\IEEEauthorblockA{Research Laboratory of Electronics\\
Massachusetts Institute of Technology\\
medard@mit.edu}}


%Telephone: (800) 555--1212, Fax: (888) 555--1212}


\maketitle

\begin{abstract}
We introduce a novel approach to error correction decoding in the presence of additive alpha-stable noise, which serves as a model of interference-limited wireless systems. In the absence of modifications to decoding algorithms, treating alpha-stable distributions as Gaussian results in significant performance loss. Building on Guessing Random Additive Noise Decoding (GRAND), we consider two approaches. The first accounts for alpha-stable noise in the evaluation of log-likelihood ratios (LLRs) that serve as input to Ordered Reliability Bits GRAND (ORBGRAND). The second builds on an ORBGRAND variant that was originally designed to account for jamming that treats outlying LLRs as erasures. This results in a hybrid error and erasure  correcting decoder that corrects errors via ORBGRAND and corrects erasures via Gaussian elimination. The block error rate (BLER) performance of both approaches are similar. Both outperform decoding assuming that the LLRs originated from Gaussian noise by $\sim$2 to $\sim$3 dB for [128,112] 5G NR CA-Polar and CRC codes. 
%Finally, an extension of our framework including the presence of jamming is proposed as well.
\end{abstract}


% \fi
\IEEEpeerreviewmaketitle

\section{Introduction}
In wireless communications, information signals can be affected by various physical phenomena: small-scale fading, additive noise due to the electronics of the transceiver, and interference \cite{goldsmith}. In order to mitigate these effects, channel coding is employed to introduce redundancy in information sequences. Error correction decoders present in the receiver chain are able recover originally transmitted sequences with high probability when sufficient redundancy has been added to overcome channel impairments \cite{5635468}. The majority of decoders proposed in the literature are characterized by two limitations: they are designed for  specific families of codes, and their performance is typically  evaluated assuming certain noise statistics, usually additive white Gaussian noise (AWGN). However, interference in wireless systems is ubiquitous and characterized by a structure that does not generally conform to AWGN \cite{7733098}. 

This paper proposes a channel decoding framework to take into account interference statistics. This framework encompasses both cases of known and unknown statistics of alpha-stable noise at the receiver using variants of Ordered Reliability Bits GRAND (ORBGRAND). The rest of this section provides an overview of the literature related to alpha-stable noise channels and GRAND.

\subsection{Alpha-Stable Noise Channels}
The studies in \cite{1145707,4802198,5560889,6064713} constitute part of the first work that considers alpha-stable distributions to model multi-user interference in wireless networks. These frameworks consider spatially distributed interferers and derive the characteristic function of the resulting aggregate interference as function of network macroparameters (e.g., base station density, transmit power, $\ldots$). The experimental studies conducted in \cite{9703600,9241843} further support the use of alpha-stable random variables to model the interference in IoT bands. These works demonstrate that received signals measured in unlicensed bands exhibit impulsive behavior since devices operating in these bands (e.g., sensors, appliances) only operate during a small fraction of time. In such cases, the tails of the observed interference distributions are heavy, making alpha-stable distributions more suitable than Gaussian models. Efforts to integrate alpha-stable models into practical communication systems are reported in \cite{6247435,5751193,6809154,9771897,8886114,8580788,7565031,5670974}, which introduce detectors and soft demappers for computing log-likelihood ratios (LLR) values associated to soft bits which serve as input to soft decision error correction decoders. Diversity combining in the presence of the same channels is studied in \cite{5529753,6399492}. In \cite{7866884}, capacity bounds for additive symmetric alpha-stable noise channels are established. 

\subsection{GRAND}
GRAND is a recently established decoder that was originally introduced for hard decision demodulation systems~\cite{8630851}.
Unlike other error correction decoders, GRAND aims to identify the binary noise effect impacting the transmission without relying on specific code structure to decode. This approach generates putative binary noise effect sequences in decreasing order of likelihood and successively tests whether what remains is a codeword when the noise effect is removed from the hard decision demodulated sequence. In hard decision settings, the query order is determined by statistical knowledge of the channel \cite{9766201}. In soft decision settings, soft input in the form of LLRs inform the query order. Soft-GRAND (SGRAND) \cite{9149208} provides a maximum likelihood decoding in the presence of soft-input. While not suitable for efficient implementation in circuits, it enables the empirical evaluation of optimal performance in the absence of computational considerations. Ordered Reliability Bits Guessing Random Additive Noise Decoding (ORBGRAND) \cite{9872126} is a computationally efficient soft decision GRAND decoder based on the principle of approximating rank ordered bit reliabilities by piecewise linear functions, which enables the
efficient generation of candidate sequences using generative integer partition algorithms. ORBGRAND has been proven to be almost capacity-achieving in AWGN channels \cite{9992258}. GRAND-EDGE and ORBGRAND-EDGE (Erasure Decoding by Gaussian Elimination) are proposed in \cite{10279273} to counteract the potential presence of jamming. In practice, hard decision GRAND and soft decision ORBGRAND can efficiently decode any moderate redundancy code of any length and are inherently well-suited to implementation in circuits due to being highly parallelizable~\cite{9567867,abbas22,other_circuit}.

\subsection{Contributions \& Notation}
Our main contributions are as follows: 
\begin{itemize}
    \item We introduce the first decoders specifically tailored to operate in alpha-stable noise channel conditions for general families of codes of moderate redundancy. 
    \item We provide multiple variants of these decoders, depending on the knowledge of the noise statistics available at the receiver.
    \item We provide an initial evaluation of performance of these decoders for a Cyclic Redundancy Check (CRC) code and a CRC-Assisted Polar (CA-Polar) code, such as is found in the 5G standard.
    %Additional performance analyses taking into account the presence of jamming are also proposed.
\end{itemize}
In the following sections, $\mathbb{F}_m$ represents the Galois Field with $m$ elements, $\vec{x}$ and $x_i$ denote a complex vector as well as its $i$th entry.
\section{Model and background}

\subsection{System model}
A binary information word $\vec{u} \in \mathbb{F}^k_2$ is encoded with a given error correcting code $\xi : \mathbb{F}^k_2 \mapsto \mathbb{F}^n_2 $ with $n>k$. The resulting code word is denoted by $\vec{c} \in \mathbb{F}^n_2$ and belongs to the code book $\mathcal{C}$ containing all the possible outputs of the encoder $\xi$. This code book is therefore defined as $\mathcal{C}=\{\vec{c}:\vec{c}=\xi(\vec{u}),\vec{u} \in \mathbb{F}^k_2\}$. Before analog transmission, constellation mapping is performed: each block of $m$ successive bits in $\vec{u}$ is mapped onto a complex symbol of a constellation. The resulting complex vector obtained after mapping is denoted by $\vec{x} \in \mathbb{C}^{n/m}$, with $m$ assumed to divide $n$. The channel introduces an additive alpha-stable noise in the transmitted symbols. The resulting signal at the receiver side is therefore defined as $\vec{Y} = \vec{x} + \vec{Z} \in \mathbb{C}^{n/m}$ where $\vec{Z}$ is a random vector whose entries follow the alpha-stable distribution recalled in the next section. 

\subsection{Alpha-stable distribution}
The alpha-stable distribution is characterized by four parameters \cite{Markov_book}: 
\begin{itemize}
    \item the stability $\alpha \in (0;2]$, characterizing the rate at which the tail of the distribution decreases;
    \item the skewness parameter $\beta \in \left[-1;1\right]$, measuring the symmetry of the distribution;
    \item the scale $\gamma \in (0;+\infty]$, characterizing the width of the density function;
    \item the location $\mu \in \mathbb{R}$, indicating where the mode of the distribution is located on the real line.
\end{itemize}
One of the challenges associated with the alpha-stable distribution is the absence of general closed-form expressions for its moments and probability density function (pdf), denoted by $f(x;\alpha,\beta,\gamma,\mu)$. An analytical expression exists for its characteristic function and is given by
\begin{equation}
\label{characteristic_function}
    \phi_{Z_i}(t) \triangleq \mathbb{E}\left[e^{jtZ_i}\right] = \exp\Big[jt\mu - |\gamma t|^\alpha \big(1-j\beta \text{sign}(t)\zeta \big)\Big]
\end{equation}
where 
\begin{equation}
  \zeta =
    \begin{cases}
      \tan(\pi \alpha /2) & \; \text{if} \; \alpha \neq 1\\
      -2 \log(|t|)/\pi & \; \text{if} \; \alpha = 1
    \end{cases}       
\end{equation}

Based on the above expression, the corresponding pdf can be retrieved using numerical inversion. 

\begin{remark}
The Gaussian distribution $\mathcal{N}(\mu,\sigma)$ can be obtained as a particular case of the alpha-stable law by setting $\alpha = 2$, $\beta=0$ and $\gamma = \sqrt{\sigma/2}$.
\end{remark}

\begin{remark}
In the framework of this study, a symmetric symbol constellation is considered. The resulting multi-user interference at symbol level is therefore symmetric as well, restricting the analysis of this paper to the parameter values $\beta = 0$ and $\mu = 0$. 
\end{remark}

Regarding the associated LLRs, no exact expression has been obtained in the literature. As mentioned in the previous section, analytical approximations have, however, been proposed. For instance, in \cite{6809154}, the approximated LLR for positive soft bits in the case of a symmetric alpha-stable distribution with Binary Phase Shift Keying (BPSK) symbols is given by
\begin{align}
    \text{LLR}(Y_i) &= \log \dfrac{f(Y_i-1;\alpha,0,\gamma,0)}{f(Y_i+1;\alpha,0,\gamma,0)} \nonumber\\
    &\approx \min \Bigg( \dfrac{\sqrt{2}}{\gamma}Y_i,2\dfrac{\alpha + 1}{Y_i} \Bigg) \label{equation_approx_LLR}
\end{align}

\begin{figure}[h!]
    \centering
    \includegraphics[width = 0.48\textwidth]{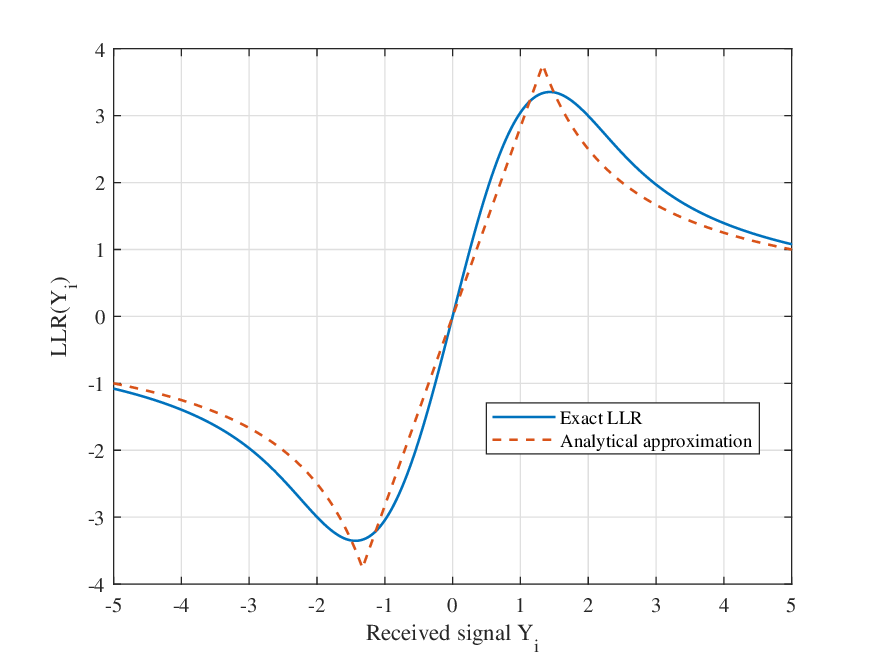}
    \caption{Numerically computed LLR values and analytical approximation from \cite{6809154}, obtained for $\alpha = 1.5$ and $\gamma = 0.5$.}
    \label{fig:LLR_approx}
\end{figure}

The accuracy of this approximation is illustrated in Fig. \ref{fig:LLR_approx}. For low values of the received signals, a behavior similar to LLR values in a Gaussian case can be observed (close to linearity). However, unlike the Gaussian case, one can note a decrease in the LLRs as the absolute value of the received soft bits further increases. Larger values of the received soft bits are therefore less and less reliable due to the heavier tails of the alpha-stable distribution.  

\section{Statistical analysis and Proposed decoders}
\label{sect:decoders}

Building upon the theoretical background of the previous section, the following paragraphs detail decoders tailored to alpha-stable noise. 

\medskip

\subsubsection{ORBGRAND-EDGE}
Originally introduced to counteract potential jamming effects \cite{10279273}, ORBGRAND-EDGE can be employed as a first solution if the parameters of the interference induced alpha-stable noise are unknown at the receiver. This method consists in erasing bits of the received signal $\vec{Y}$ which are characterized by extremely high soft values $Y_i$. These outliers are, indeed, likely to arise from the impulsive nature of the interference, which features heavier tails than the Gaussian distribution. In the case of alpha-stable noise, these extreme soft values are characterized by a decreasing reliability (see LLR in Fig. \ref{fig:LLR_approx}) motivating their erasure. A threshold $\delta$ is therefore defined such that bits satisfying $|Y_i|>\delta$ are treated as erasures. 

Once bits meeting that condition are removed, classical ORBGRAND is applied to the remaining elements of the codeword, leveraging the ability of GRAND algorithms to decode any code structure. In order to establish a list of candidate noise sequences, the LLR values associated to these remaining bits should be computed and provided to ORBGRAND. Assuming that the distribution of the alpha-stable noise $f(x;\alpha,\beta,\gamma,\mu)$ is unavailable at the receiver, the exact computation of these values is not possible. To circumvent this issue, the LLRs are instead computed by treating the noise affecting remaining bits as AWGN. This approximation is justified by the tendency of the alpha-stable LLR curve which is, for low values of $Y_i$, close to the linear behavior of LLRs in the presence of AWGN noise (see Fig. \ref{fig:LLR_approx}). When a candidate sequence is proposed by ORBGRAND, the algorithms attempts to recover the original hard bits associated to the erased elements using Gaussian elimination. If, for a given codeword, no unique solution can be obtained from the resulting linear system, the original ORBGRAND is applied on the whole block (without erasing any bits).
\begin{remark}
\label{optimal_thresh_edge}
It is relevant in this context to fine-tune this decoder with respect to its erasure threshold $\delta$. Figure \ref{fig:sensitivity_delta} illustrates the bit error rate (BER) obtained with ORBGRAND-EDGE as function $\delta$. These results have been obtained for a [128,112] cyclic redundancy check (CRC) code, $\alpha = 1$ and values of $\gamma$ leading to the indicated SNRs. The results suggest the existence of optimum thresholds minimizing the BER for each considered SNR. The presence of these minima can be justified by the following arguments: for excessively high values of $\delta$, the algorithm will not erase some of the outliers, which might compromise the decoding process. In contrast, for very low values of the threshold, the algorithm might erase too many elements of the code word, resulting in a linear system in the Gaussian elimination that will not admit a unique solution. In that case, the algorithm will by default resort to applying ORBGRAND on the whole block (without any erasures). The asymptotic performance of ORBGRAND-EDGE for both $\delta \rightarrow 0$ and $\delta \rightarrow +\infty$ should therefore be identical to the classical ORBGRAND decoder. The intermediate minimum therefore corresponds to the optimum mean proportion of elements to be erased within the block.
\begin{figure}[h!]
    \centering
    \includegraphics[width = 0.48\textwidth]{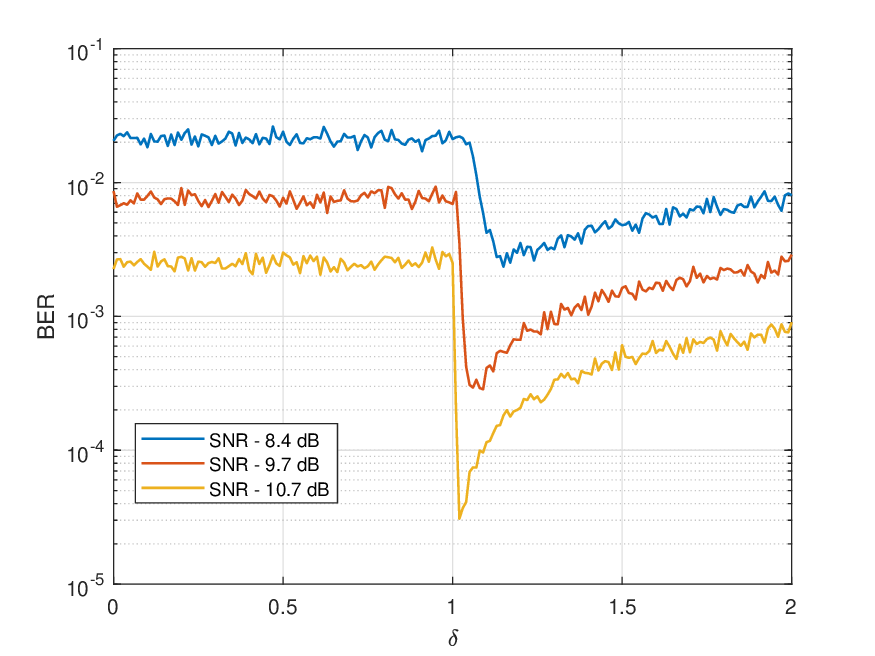}
    \caption{Sensitivity analysis of ORBGRAND-EDGE with respect to the threshold $\delta$.}
    \label{fig:sensitivity_delta}
\end{figure}
\end{remark}
\subsubsection{$\alpha$-ORBGRAND-EDGE (based on reliabilities)}
Now assuming that the parameters of the alpha-stable pdf of the noise are available at the receiver, this decoder eliminates elements of the received codewords using their associated reliability values $\text{LLR}(Y_i)$ instead of their soft values $Y_i$. This method requires another threshold $\epsilon$ chosen such that bits satisfying $|\text{LLR}(Y_i)|<\epsilon$ are erased. From Fig. \ref{fig:LLR_approx}, one can deduce that this algorithm eliminates bits with extremely high values $Y_i$ (similarly to ORBGRAND-EDGE), but also bits of values $Y_i$ close to zero (i.e., near the boundaries of the BPSK constellation zones). The remaining bits are decoded using $\alpha$-ORBGRAND which is also based on $\text{LLR}(Y_i)$ (see below). In case no unique solution can be found to the linear system associated to the Gaussian elimination, the same algorithm is employed to decode the whole block without erasure.

\subsubsection{$\alpha$-ORBGRAND}
This decoder generalizes ORBGRAND methods by incorporating the exact nonlinear LLR values of the noise, computed from their alpha-stable pdf. Once they have been computed, these LLRs are sorted in ascending order. Linearly approximating these sorted LLR when expressed as function of their rank order enables to efficiently generate candidate noise sequences \cite{9872126}.

\subsubsection{$\alpha$-SGRAND}
The list of candidate noise sequences generated by this new decoder also relies on LLRs computed from the alpha-stable noise distribution. However, unlike $\alpha$-ORBGRAND, this list is dynamic: at each time step, the most probable candidate is tested, and the list is updated with new putative sequences. The execution of the algorithm based that list is shown in  \cite{9149208} to cover vectors in a non-increasing order of likelihood. The returned solution is therefore proven to be a maximum likelihood code word.

\begin{remark}
    The three latter methods require computing the LLRs of the received soft bits. This computation theoretically requires the knowledge of the corresponding alpha stable pdf at the receiver side. As no closed-form expression exists for this pdf, its values must either be stored in memory in advance, or obtained via numerical inversion of \eqref{characteristic_function}. For this reason, the approximation of \eqref{equation_approx_LLR} might be preferable. In both cases, the estimation of parameters $\alpha$ and $\gamma$ of the distribution is required. Estimation algorithms for these parameters based on data samples have been proposed in \cite{estimation_1} and \cite{estimation_2}. 
    %The impact of potential estimation errors on the performance is discussed in the next section.
\end{remark}
\section{Numerical results}
\label{sect:numerical results}

\begin{figure}[h!]
    \centering
    \includegraphics[width = 0.48\textwidth]{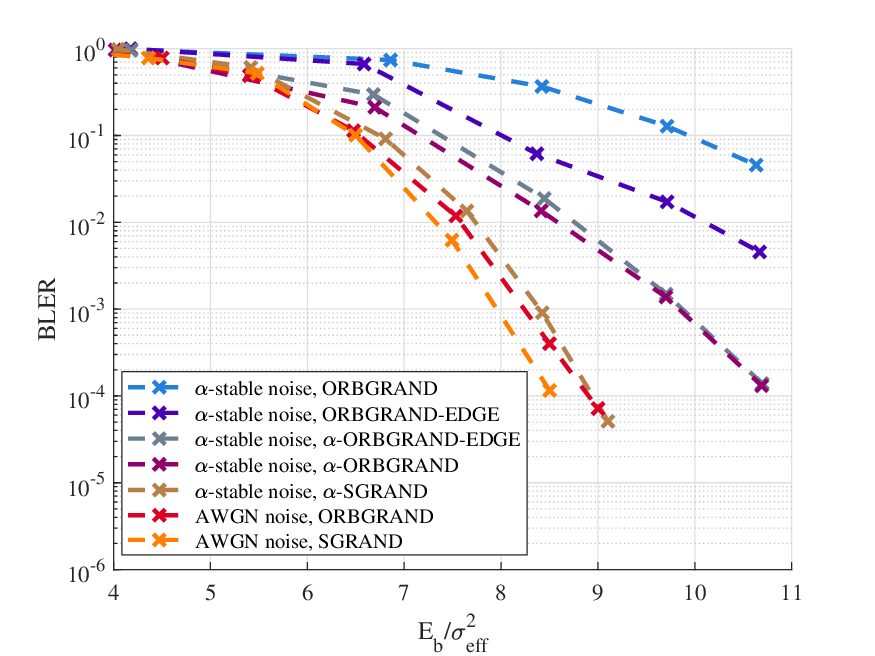}
    \caption{Decoding performance obtained for a [128,112] CRC code in additive alpha-stable noise.}
    \label{fig:perf_CRC}
\end{figure}

\begin{figure}[h!]
    \centering
    \includegraphics[width = 0.48\textwidth]{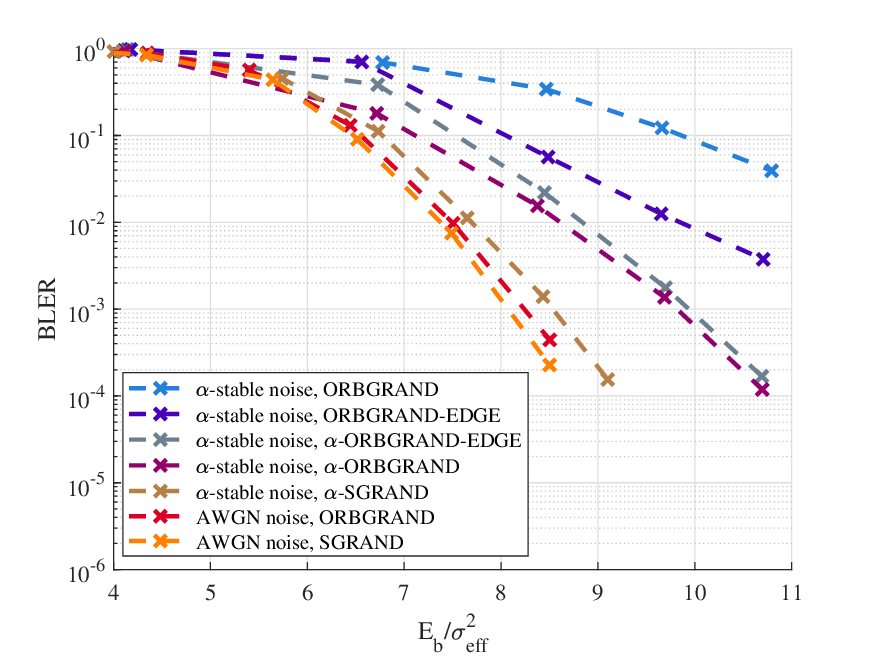}
    \caption{Decoding performance obtained with a [128,112] CA-Polar code in additive alpha-stable noise.}
    \label{fig:perf_CAPOLAR}
\end{figure}

The performance of the decoders is illustrated in Figs. \ref{fig:perf_CRC} and \ref{fig:perf_CAPOLAR}. Two families of codes have been employed to generate these graphs. These families have been shown in \cite{9500279,9086275} to provide competitive performance with GRAND algorithms. A CRC code has been used for Fig. \ref{fig:perf_CRC}. This family was primarily used for error detection. The rationale behind its use for error correction and its correcting capabilities with GRAND are detailed in \cite{9500279}. Fig. \ref{fig:perf_CAPOLAR} has been produced using a CRC-assisted polar (CA-Polar) code. These codes are employed for control channel communications in 5G New Radio. Their performance with GRAND decoders is demonstrated in \cite{9086275}. For consistency, the code dimensions $[n,k]$ have been set to $[128,112]$ for both figures. The corresponding level of complexity allows for the production of the results of this section using simple versions of the algorithms, without the need for parallelization. The CRC length is given by $\ell = 11$ for the CA-Polar code. Regarding the parameters of the alpha-stable distribution, $\alpha$ is set to 1 and $\gamma$ varies to produce the equivalent\footnote{In the case of Gaussian noise, the variance of the normal distribution directly relates to the noise power $N_0$ present in the denominator of the SNR $E_b/N_0$, where $E_b$ is the energy per bit. For a BPSK constellation, the corresponding error probability $p_e$, obtained with a hard demodulator before decoding, can be analytically deduced using the inverse complementary error function. In the presence of alpha-stable noise, such a direct connection between variance and SNR ratio does not exist since the moments of the alpha stable distribution are undefined. Consequently, performance results of decoders in the presence of alpha-stable and Gaussian noises are compared for \textit{equivalent} SNR, or in other words, for an equivalent error probability $p_e$.} SNR $E_b/\sigma_{\text{eff}}^2$ on the horizontal axis of the graphs.

\smallskip

The block error rate (BLER) obtained if instead of using the alpha-stable distribution one assumes AWGN is shown in light blue curves in both Figs. \ref{fig:perf_CRC} and \ref{fig:perf_CAPOLAR}. Code words are in that case decoded using  ORBGRAND, and the associated LLRs are computed by treating the noise as AWGN. A first gain in performance is illustrated in dark blue if one uses ORBGRAND-EDGE with a fixed $\delta = 1.2$. Following Remark \ref{optimal_thresh_edge}, this gain could be further enhanced using erasure thresholds tuned for every SNR. This improvement is left for future work.

Assuming that the noise distribution is known at the receiver, the performance can be further improved using $\alpha$-ORBGRAND-EDGE (illustrated in grey for $\epsilon = 3$). As explained in the previous section, this algorithm can eliminate a larger number of problematic bits than ORBGRAND-EDGE since its erasure criterion is based on LLR values. The similar performance obtained with $\alpha$-ORBGRAND (purple curves) can be explained by two reasons. First, even though their values are low, the LLRs associated to the bits erased using $\alpha$-ORBGRAND-EDGE are nonzero and still contain a certain amount of information. As a result, the associated soft information could still be taken into account in the rank ordering performed by $\alpha$-ORBGRAND. Second, $\alpha$-ORBGRAND-EDGE strongly relies on $\alpha$-ORBGRAND to decode unerased bits, and when the Gaussian elimination is not possible. These results suggest that $\alpha$-ORBGRAND-EDGE could be further studied for longer codes (with higher redundancies) and scenarios featuring jamming or more interference.

The best BLER in the presence of alpha-stable noise is obtained using $\alpha$-SGRAND (in beige) since this decoder utilizes soft information to produce maximum likelihood code words. The BLER with AWGN noise, which features no outlier, is also illustrated for comparison when using classical ORBGRAND and SGRAND (red and orange curves). It is notable that the best BLER under alpha-stable conditions is slightly higher but remains close (within 1dB) to the results obtained with Gaussian noise and SGRAND. Note that Gaussian noise is, for given energy, pessimal in terms of capacity. 

\section{Conclusion}
In this paper, we present  the first decoders designed to operated in alpha-stable noise conditions, for general moderate redundancy codes. We have assessed the performance of these decoders in some initial settings.

\begin{figure}[h!]
    \centering
    \includegraphics[width = 0.48\textwidth]{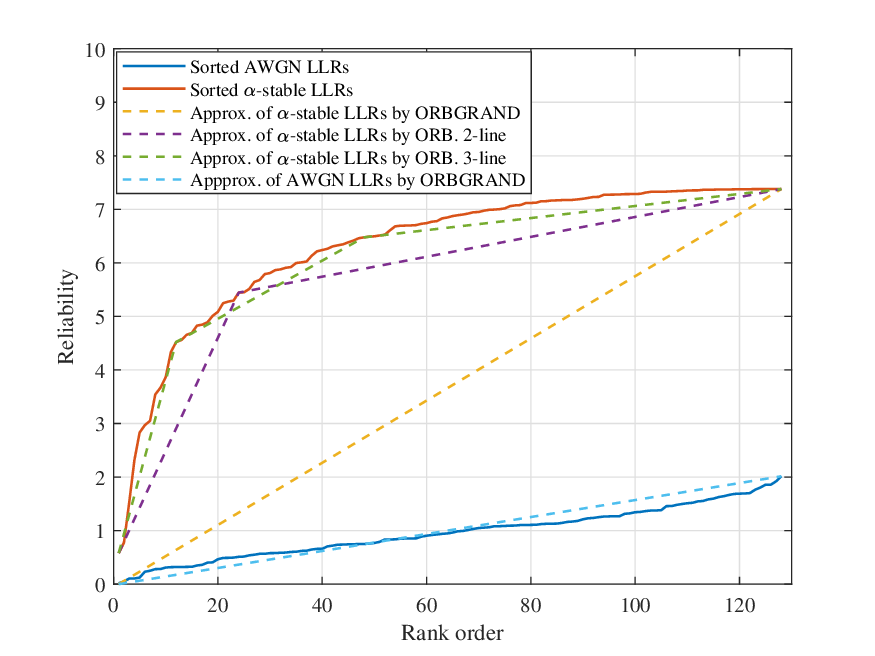}
    \caption{Rank ordered LLR values associated to the bits of a codeword affected by Gaussian noise (continuous blue line) and by $\alpha$-stable noise (continuous red line). These results have been generated for a BPSK constellation, a block of size 128, a symmetric alpha stable noise of parameters $(\alpha,\gamma) = (1,0.05)$, and a Gaussian noise of equivalent SNR. Approximations of these reliabilities obtained by means of OBRGRAND and its multiple line variants are represented in dashed lines.}
    \label{fig:sorted_LLR}
\end{figure}

Future work includes developing multiple line versions of $\alpha$-ORBGRAND to better approximate rank ordered LLRs in the presence of alpha-stable noise. As shown in Fig. \ref{fig:sorted_LLR}, the linear approximation of ORBGRAND, used to generate noise sequence candidate lists, tends to accurately approximate sorted reliability values in the AWGN case. By contrast, resorting to linear piecewise approximations (two or three lines) as shown in Figure \ref{fig:sorted_LLR} would be more accurate in alpha-stable noise conditions.   

Our work considered a simple BPSK modulation and interference with short moderate-redundancy codes, but no multiple access or channel fading.  GRAND has also been applied to assist in optimal modulation \cite{10001177} and  multiple access channels \cite{9517890,10356355}. While we considered noise statistics, a natural  additional consideration is that of   fading channels, which can be taken into account with GRAND \cite{10001707,10008698,9766201,10437763}. While we considered short codes with moderate redundancy, GRAND can also be applied to long and low-rate codes (e.g., product codes) \cite{9685645}. Bringing interference through alpha-stable and related noise distributions to such settings opens a wide array of future research.

\section*{Acknowledgment}
This work was supported by the Defense Advanced Research Projects Agency (DARPA) under Grant HR00112120008.

\bibliographystyle{IEEEtran}
\bibliography{bibli}

\end{document}